\documentclass{iopart}                
\usepackage{graphicx}
\usepackage{amssymb}
\newcommand{\bra}[1]{\langle {#1} |}
\newcommand{\ket}[1]{| {#1} \rangle}

\newcommand{\Ham}{\mathcal{ H}}

\begin{document}

\letter{A basis of cranking operators
for the pairing-plus-quadrupole model}
\author{Takashi~Nakatsukasa\dag%
\footnote[3]{Electronic address: {\tt T.Nakatsukasa@umist.ac.uk}}%
,
Niels~R.~Walet\dag%
\footnote[4]{Electronic address: {\tt Niels.Walet@umist.ac.uk}}%
, and Giu~Do~Dang\ddag%
\footnote[5]{Electronic address: {\tt Giu.Dodang@th.u-psud.fr}}%
}
\address{\dag Department of Physics, UMIST, P.O.Box 88,
Manchester M60 1QD, UK}
\address{\ddag Laboratoire de Physique Th\'eorique et Hautes Energies,\\
B\^at 211, Universit\'e de Paris-Sud, 91405 Orsay, France}

\begin{abstract}
We investigate the RPA normal-mode coordinates in the
pairing-plus-quadrupole model, with an eye on simplifying the
application of large amplitude collective motion techniques.
At the Hartree-Bogoliubov minimum, the RPA modes are exactly the
cranking operators of the collective coordinate approach.
We examine the possibility of representing the self-consistent cranking
operator by linear combinations of a limited number of one-body
operators.
We study the Sm nuclei as an example, and find
 that such representations exist in terms of operators
that are state-dependent in a characteristic manner.
\end{abstract}


The selection of proper collective variables is an important problem
in the study of large amplitude collective motion.  In the usual
Constrained Hartree-Fock (CHF) or Hartree-Fock-Bogoliubov (CHFB)
calculations, the collective subspaces are generated by a small number
of one-body constraint (also called cranking) operators which are most
commonly taken to be of the multipole form ($r^L Y_{LK}$). In
realistic calculations of processes such as fission, the number of
coordinates to describe the full nuclear dynamics can easily become
larger than can be dealt with in satisfactory manner, and a method to
determine the optimal combination needs to be devised. Even assuming
that such a method exists, there is no {\it a priori} reason to limit
oneself to multipole operators, and the cranking operators should be
determined by the nuclear collective dynamics itself, from the set
of all one-body operators. This is clearly a difficult task, and one
would like to be able to select a small group of operators, and
find the optimal combination of these operators at each point of the
collective surface.

In our past work, we have investigated a theory of adiabatic large
amplitude collective motion as a method to generate self-consistent
collective subspaces (see reference ~\cite{KWD91} and references
therein). The key ingredient of the method is the self-consistent
determination of the constraint operator, and as such it may provide an
answer to the selection question discussed above.  Using the local
harmonic version (LHA) of the
theory \cite{KWD91}, we have recently embarked on a study of the
properties of large amplitude collective motion in systems with
pairing.  We have dealt with two simple models: 
a semi-microscopic model of nucleons interacting through a pairing
force, coupled to a single harmonic variable \cite{NW98} and
 a fully microscopic $O(4)$ model which may be regarded as a
simplified version of the pairing-plus-quadrupole (P+Q) Hamiltonian
\cite{NW99}.  It has turned out that the self-consistent collective
coordinate obtained by the LHA accounts quite well for the exact
dynamics of these models.  We have also shown that the CHFB calculations
using the mass-quadrupole operator as the constraint operator can
result in incorrect results \cite{NW99}.  It is not immediately
obvious that we can extrapolate the conclusions reached in these
models to fully realistic nuclear problems.  Within the time-dependent
Hartree-Bogoliubov (TDHB) approximation, the dynamics of these models
is described by  a modest number of degrees of freedom ($4\sim
12$).  For realistic problems in heavy nuclei, on the other hand, we
need to deal with millions of degrees of freedom!  In this letter, we
report the first attempts to study such realistic nuclear problems.

In the LHA, the collective path (or collective surface for more than
one coordinate) is determined by solving the CHFB problem with a
cranking operator which is self-consistently determined by the local
RPA.  Since this procedure requires us to solve the RPA at each point
on the collective path, it will be very useful if the RPA
eigenvectors can be approximated by taking linear combinations of a
small number of one-body operators. This means that we can restrict
the RPA diagonalisation to this small space, rather than deal with
the full millions-by-millions RPA matrix.  Such a scenario has been
examined in reference ~\cite{PL77} and in reference ~\cite{DWK94} for the HF
problem in $^{28}$Si, and a strong radial and spin dependence of
self-consistent cranking operators was found.  However, since the
model space had only six degrees of freedom and the pairing degrees of
freedom were neglected, the results can not be directly generalised
to heavy nuclei.  Here we perform a similar analysis for
the P+Q model.  Since
the model is known to be able to realistically describe collective
phenomena involving both pairing and quadrupole degrees of freedom
\cite{KS60,BK65}, we expect that the same choice for the set of
one-body operators should work for other realistic Hamiltonians as well.

Baranger and Kumar analysed in great detail the collective
motion in the P+Q model assuming that the collective variables are the
mass quadrupole operators \cite{BK65}.  Thus, they reduced the large
number of two-quasiparticle (2qp) degrees of freedom (of the order of
a thousand) into only two ``collective'' coordinates, $\beta$ and
$\gamma$.  However, our previous study of the $O(4)$ model \cite{NW99}
suggests that even for such simple Hamiltonians the self-consistent
collective coordinate is not as trivial as it seems to be.  In this
letter we report the first result for the P+Q model and show that the
normal-mode coordinate of the random-phase approximation (RPA) is
quite different from the mass quadrupole operator.  This is especially
true when the system is deformed.

For the P+Q model, the Hartree-Bogoliubov (HB) ground state can be
specified by 6 parameters ($\Delta_n$, $\Delta_p$, $\lambda_n$,
$\lambda_p$, $\beta$ and $\gamma$).  The TDHB equations can be
shown to be equivalent to Hamilton's equations of motion \cite{BR86}.
The underlying classical Hamiltonian can be written, up to second
order in coordinates and momenta, as
\begin{equation}
\Ham \equiv \bra{\Psi} H-\mu N \ket{\Psi} \approx
 E_0 + \frac{1}{2} B^{\alpha\beta} \pi_\alpha \pi_\beta
               + \frac{1}{2} C_{\alpha\beta} \xi^\alpha \xi^\beta \ ,
\label{H_HA}
\end{equation}
in terms of the canonical variables $(\xi,\pi)$ \cite{BR86}.
Here each of the indices ($\alpha,\beta,\cdots$) indicates a pair of
2qp indices ($ij,kl,\cdots$).
We adopt the standard tensor notation where a repeated upper and lower index
indicates a summation.
The mass and curvature parameters are 
\begin{equation}
\fl B^{\alpha\beta}      = 
 E_\alpha \delta_{\alpha\beta}
   - 2\sum_\rho \chi_\rho S^{(\rho)}_\alpha {S^{(\rho)}_\beta} \ , 
\qquad
C_{\alpha\beta}        = 
 E_\alpha \delta_{\alpha\beta}
   - 2\sum_\rho \chi_\rho R^{(\rho)}_\alpha {R^{(\rho)}_\beta} \ .
\end{equation}
Here $R^{(\rho)}$ and $S^{(\rho)}$ represent the hermitian and
anti-hermitean components, respectively, of the pairing
and quadrupole operators,
\begin{equation}
\fl R^{(\rho)} = \sum_\alpha R^{(\rho)}_\alpha \left(
            (a^\dagger a^\dagger)_\alpha + \mbox{h.c.} \right) \ ,
\qquad
S^{(\rho)} = \sum_\alpha S^{(\rho)}_\alpha \left(
            (a^\dagger a^\dagger)_\alpha - \mbox{h.c.} \right) \ ,
\label{2qp_matrix_elements}
\end{equation}
where the scattering terms $(a^\dagger a)$ are omitted.
Following reference ~\cite{BK65},
we multiply the quadrupole operators by a factor $\alpha_\tau^2$
with $\alpha_n=(2N/A)^{1/3}$ and $\alpha_p=(2Z/A)^{1/3}$,
and also reduce
the quadrupole matrix elements between the states of the upper
shell by a factor
$\zeta = (\mathcal{ N}_{\rm L}+\frac{3}{2})/(\mathcal{ N}+\frac{3}{2})$,
where $\mathcal{ N}$ is the oscillator quantum number operator and
$\mathcal{ N}_{\rm L}$ is the number of quanta in the lower shell.
Thus, the modified quadrupole operators are defined as
$Q_{2K} \equiv (Q_{2K})_n + (Q_{2K})_p$, with
$(Q_{2K})_n=\alpha_n^2\zeta (r^2 Y_{2K})_n$
and $(Q_{2K})_p=\alpha_p^2\zeta (r^2 Y_{2K})_p$
(which  we shall refer to as ``the quadrupole operators'').
The RPA equation is solved by diagonalisation of the Hamiltonian
(\ref{H_HA}). This can be implemented by a (linear) transformation to 
normal coordinates
($q^\mu$),
\begin{equation}
q^\mu =  f^\mu_{,\alpha} \xi^\alpha, \quad\quad
\xi^\alpha = g^\alpha_{,\mu} q^\mu  \ ,
\end{equation}
where $f^\mu_{,\alpha}$ and $g^\alpha_{,\mu}$ are just coefficients of
linear transformation.  These can be also identified as $\partial
q^\mu/\partial \xi^\alpha$ and $\partial\xi^\alpha/\partial q^\mu$ for
the general point transformation, $q^\mu=f^\mu (\xi)$ and
$\xi^\alpha=g^\alpha(q)$, when we study large
excursions from  equilibrium \cite{KWD91}.  In the
case of a single collective coordinate, the path is determined by
solving the CHB equation with the cranking operator $q^1=f^1(\xi)$.  A
similar set of equations has been also obtained by other adiabatic
time-dependent mean-field theories \cite{RS80}.

The solution of the RPA equation
\begin{equation}
C_{\alpha\gamma}B^{\gamma\beta} f^\mu_{,\beta}
 = (\Omega^\mu)^2 f^\mu_{,\alpha} \ ,
\label{full_RPA_eq}
\end{equation}
involves the diagonalisation of the RPA matrix $C_{\alpha\gamma}B^{\gamma\beta}$
whose dimension is equal to the number of active 2qp degrees of
freedom.
For separable forces, this can be simplified by solving a dispersion relation.
In general, however, the RPA diagonalisation requires extensive computational
resources.
Now let us approximate an eigenvector using a certain set of one-body operators
$\{ O^{(i)} \}$:
\begin{equation}
\bar{f}_{,\alpha} = \sum_i C_i O^{(i)}_\alpha \ ,
\end{equation}
where $O^{(i)}_\alpha$ indicate the 2qp matrix elements of operator $O^{(i)}$
as in equation~(\ref{2qp_matrix_elements}).
Then, instead of the full RPA equation (\ref{full_RPA_eq}),
we obtain a  projected RPA equation
\begin{equation}
\bar{M}^{ij} C^n_j = \left( \bar{\Omega}^n \right)^2
                           \bar{N}^{ij} C^n_j \ ,
\label{projected_RPA_eq}
\end{equation}
which  determines the coefficients $C^n_i$, with
\begin{equation}
\bar{M}^{ij}\equiv 
O^{(i)}_\alpha B^{\alpha\beta} C_{\beta\gamma}
                             B^{\gamma\delta} O^{(j)}_\delta \ ,
\quad\quad
\bar{N}^{ij} \equiv O^{(i)}_\alpha B^{\alpha\beta} O^{(j)}_\beta \ .
\label{projected_RPA_matrix}
\end{equation}
The dimension of the matrices $\bar{M}^{ij}$ and
$\bar{N}^{ij}$ is equal to the number of one-body
operators $\{O^{(i)}\}$.  Therefore, if we can approximate the RPA
eigenvectors by using a small number of operators, it will
significantly reduce the computational task.

A criterion for good projection may be given by the closeness of the
projected RPA frequencies $\bar{\Omega}$ to the real
RPA frequencies $\Omega$.
Another criterion is the smallness of the quantity $\delta$,
\begin{equation}
\label{delta}
\delta = f_{,\alpha} B^{\alpha\beta}
         \left( f_{,\beta} - \bar{f}_{,\beta} \right) \ .
\end{equation}
If we normalise $f^\mu$ as
$f^\mu_{,\alpha} B^{\alpha\beta} f^\nu_{,\beta} = \delta^{\mu\nu}$,
we find $0 \leq \delta \leq 1$ with
$\delta=0$ corresponding to the exact projection and $\delta=1$ to the
case where $\bar{f}$ is orthogonal to $f$.

We have performed calculations for several heavy isotopes. Here we
report the numerical results for even-even Sm isotopes
(A=146$\sim$154).  The form of the  P+Q model is that discussed in the
second and third of the series of papers by Baranger and Kumar \cite{BK65}.
The model space and the parameters, such as the spherical
single-particle energies, the pairing and quadrupole force strengths,
are taken from table 1 in the third paper.  The equilibrium
parameters ($\beta$, $\gamma$, $\Delta$, $\lambda$) are found to agree
with table 2 of the same paper.  The ground states of $^{146,148}$Sm
are spherical ($\beta=0$) and the others have  prolate shapes
($\beta>0, \gamma=0$).

\begin{figure}
\begin{indented}
\item[]\includegraphics[width=10cm]{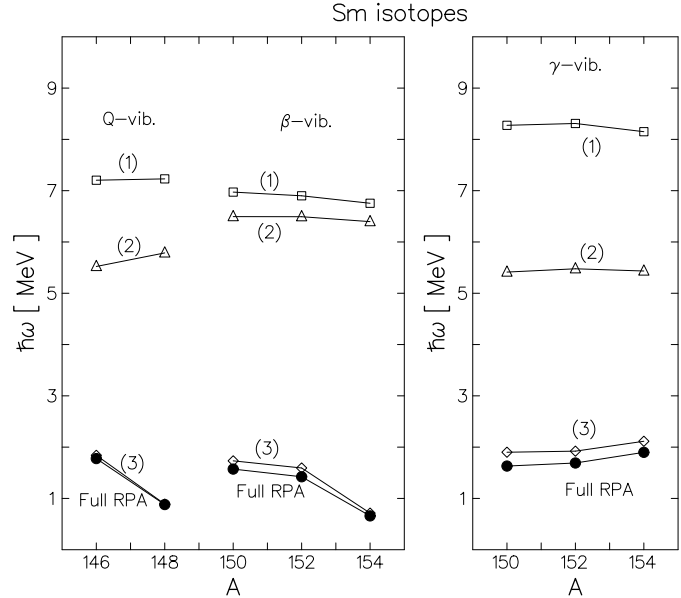}
\end{indented}
\caption{Calculated excitation energies of quadrupole, $\beta$ and
$\gamma$ vibrations for even-even Sm isotopes.
Note that the ground states of $^{146,148}$Sm are spherical.
The closed circles indicate the RPA results while the open symbols 
are the results of projected RPA calculations.
See the main text for the difference between (1), (2) and (3).
}
\label{fig:sm_omega}
\end{figure}

Figure~\ref{fig:sm_omega} shows the excitation energies (RPA
frequencies) of $\beta$ and $\gamma$ vibrations, obtained by the RPA
and projected RPA calculations.  For the projected RPA calculations,
we have adopted three different sets of one-body operators.  The first
and simplest choice is to use the operators appearing in the separable
forces, the pairing and quadrupole operators, $P_\tau$,
$P_\tau^\dagger$, $(Q_{2K})_\tau$ ($\tau=n,p$).  This choice is denoted
as (1) in the figure.  In this case the projected RPA matrices of
equation~(\ref{projected_RPA_matrix}) are two dimensional for spherical
nuclei and for $\gamma$ vibrations, while they are six dimensional for
the $\beta$ vibrations.  The calculated frequencies are $7\sim 8$ MeV
which are $5\sim 6$ MeV larger than the corresponding RPA
frequencies.  In the second set, labelled as (2), we increase the
number of operators. We keep both the pairing and quadrupole operators,
but include two additional quadrupole operators with ``monopole'' radial
dependence, $(r^0Y_{2K})_\tau$. We also include the
hexadecapole operators, $(r^4 Y_{4K})_\tau$, and the rank-2
spin-dependent operators, $([r^0Y_2 \times {\bf s} ]^{(2)}_K)_\tau$,
$([r^2Y_2 \times {\bf s} ]^{(2)}_K)_\tau$.  As far as the frequencies
are concerned, we can see some improvement over the case (1) for
spherical and the $\gamma$ vibrations in deformed nuclei, though they
are still much higher than the real RPA frequencies.  For the $\beta$
vibrations, the inclusion of the additional rank-2 (and higher rank)
operators seems not so important.  Actually we see that the $\beta$
vibrations are found to have a significant amount of monopole
components.  For the last set, denoted as (3), we adopt the same operators
as (1) but each 2qp matrix element is weighted with a factor $(E_{\rm
2qp})^{-2}$.  This means that we employ a set of {\it state-dependent}
one-body operators $\{ \widetilde{O}^{(i)}\}$ defined by
\begin{equation}
\widetilde{O} \equiv \sum_\alpha
    \frac{O_\alpha}{(E_\alpha)^2} (a^\dagger a^\dagger)_\alpha
    + \mbox{h.c.} \ .
\label{scale}
\end{equation}
The result of this projection is now almost identical to that of the full RPA.

\fulltable{Calculated values of $\delta$, equation~(\ref{delta}),
for the projected RPA solutions for Sm isotopes.
The columns (1), (2), (3), (1-a) and (1-b)
represent the different projections (see text).
For the spherical nuclei ($^{146,148}$Sm),
there is no distinction between $\beta$ and $\gamma$ vibrations.}
\label{values_of_delta}
\setlength\tabcolsep{5.5pt}
\begin{tabular}{@{}lllllllllll}
\br
A & \centre{5}{$\beta$ vibration} &
   \centre{5}{$\gamma$ vibration} \\\ns
& \crule{5} & \crule{5}\\
 & (1) & (1-a) & (1-b) & (2) & (3)  & (1) & (1-a) & (1-b) & (2) & (3)\\ 
\mr
146 & 0.271 & 0.132 & 0.225 & 0.421 & 0.009 & & & & &\\
148 & 0.243 & 0.131 & 0.184 & 0.314 & 0.0003 & & & & &\\
150 & 0.602 & 0.499 & 0.519 & 0.632 & 0.026
    & 0.610 & 0.342 & 0.507 & 0.685 & 0.092\\
152 & 0.497 & 0.346 & 0.433 & 0.526 & 0.020
    & 0.616 & 0.279 & 0.472 & 0.691 & 0.081\\
154 & 0.513 & 0.117 & 0.437 & 0.534 & 0.002
    & 0.636 & 0.208 & 0.426 & 0.679 & 0.052\\ \br
\end{tabular}
\endfulltable

In table~\ref{values_of_delta}, the quality of projection $\delta$,
equation~(\ref{delta}), is listed.  In the cases (1) and (2), where the RPA
vectors are projected on the elementary operators, $\delta \gtrsim
0.25$ for $^{146,148}$Sm and $\delta \gtrsim 0.5$ for the others.
Therefore, roughly speaking, the one-body operators possess at
most 75\% (50\%) of overlap with the real eigenvectors in spherical
(deformed) nuclei.  On the other hand, the projection (3) exhausts
more than 90\% of real eigenvectors even for the worst case.  At first sight
it may
look strange that $\delta$ is larger for (2) than for (1), while the energy
for (2) is lower.  This is due to the fact that case (2)
is dominated by certain neutron components.  Since the relevant
neutron 2qp energies are lower than those of protons, this
proton-neutron asymmetry leads to a decrease in the frequency
$\bar{\Omega}$ and at the same time an increase in $\delta$. This is
also a reflection of the poor quality of the approximation.

The figure and table indicate
that it is very difficult to obtain sensible results
by using elementary one-body operators (i.e., of the form (1) or (2)).
This is mainly due to the fact that the RPA eigenvectors, when being projected
onto elementary one-body operators, have
unrealistically large amplitudes for high-lying 2qp components.
In order to demonstrate this, we introduce a cut-off energy $\Lambda_{\rm cut}$
for the 2qp matrix elements,
i.e., $O^{(i)}_\alpha = 0$ for $E_\alpha > \Lambda_{\rm cut}$.
We then perform the projected RPA calculation
with the truncated one-body operators from set (1).
The resulting values  $\delta$ are listed in table~\ref{values_of_delta}
for $\Lambda_{\rm cut}=5$ MeV (1-a) and for 10 MeV (1-b).
We see that the major contributions to the RPA modes come
from the 2qp components
with $E_{\rm 2qp} < 5$ MeV. We thus conclude that
the superiority of the projection (3) simply comes from its being capable of
suppressing
the unnecessary high-energy components by the factor $(E_{\rm 2qp})^{-2}$.
This suppression factor is not arbitrary, but can be derived
from the following simple argument.
If we have a single-mode separable force $H=-(1/2)\chi R R$
(assuming no coupling among different modes),
we can analytically determine the RPA eigenvectors,
$f^\mu_{,\alpha} \propto R_\alpha/((E_\alpha)^2-(\Omega^\mu)^2)$.
In the limit that $\Omega^\mu \ll E_{\rm 2qp}$,
the projection on $f_{,\alpha} \propto R_\alpha/(E_\alpha)^2$
gives the exact answer.

Let us examine the projection (3) in more  detail.  For spherical Sm
nuclei, the RPA eigenvector is of isoscalar character and can be
approximated as $\bar{f} \approx (\widetilde{Q}_2)_n +
(\widetilde{Q}_2)_p$ where the tilde indicates that the matrix elements
include the suppression factor as in equation (\ref{scale}).
For deformed nuclei, where the collectivity of the
vibrational states is smaller than for spherical nuclei and the pairing
modes can mix with the quadrupole ones, the situation is more complex.
Taking $^{154}$Sm as an example, the
eigenvectors of $\beta$ and $\gamma$ vibrations are
\begin{eqnarray}
\fl \bar{f}^{\beta{\rm-vib}}&=& (\widetilde{Q}_{20})_n
 + 0.91 (\widetilde{Q}_{20})_p 
 - 0.48 \widetilde{P}_n - 0.44 \widetilde{P}_p
 + 0.085 \widetilde{P}_n^\dagger - 0.14 \widetilde{P}_p^\dagger\ ,\\
\fl \bar{f}^{\gamma{\rm-vib}}&=& (\widetilde{Q}_{22})_n
   + 0.87 (\widetilde{Q}_{22})_p \ .
\end{eqnarray}
For the $\beta$ vibration, we find a significant mixing  with the
monopole pairing modes.

In conclusion, we have examined the possibility of expressing
the self-consistent
cranking operator in terms of a limited set of one-body operators.
It seems very difficult to approximate the normal-mode vectors
with use of elementary one-body operators. This difficulty disappears, however,
when we use a small number of {\it state-dependent} one-body operators.
This may reflect the importance of the self-consistent determination of
the collective coordinates for large amplitude collective motion,
because the coordinates now have a strong state-dependence as well.
The structure of the self-consistent cranking operators is clearly
changing when we move from spherical to deformed nuclei.
For the study of  large amplitude
collective motion in heavy nuclei for which
the diagonalisation of the RPA matrix becomes too time-consuming,
the results of this paper may give a hint for a correct
choice of a state-dependent basis of operators.
The choice of a limited set of (state dependent) basis operators
provides a practical way to solve the LHA through the projection.
With the self-consistent cranking operators,
the LHA should provide a significant improvement over the conventional
CHFB calculation based on  fixed cranking operators.
Clearly we have not discussed the  structure of the self-consistent 
cranking operator away from the minimum point. 
This will be the subject of a future publication.

This work was supported by a research grant (GR/L22331) from
the Engineering and Physical Sciences Research Council (EPSRC)
of Great Britain, and through a grant (PN 98.044) from
Alliance, the Franco-British Joint Research programme.
The Laboratoire de Physique Th\'eorique et Hautes Energies is a 
Laboratoire associ\'e au C.N.R.S., URA D00063.



\end{document}